# Micromirror total internal reflection microscopy for high-performance single particle tracking at interfaces

Xuanhui Meng[1,#], Adar Sonn-Segev[1,2,#], Anne Schumacher[1], Daniel Cole[1,2], Gavin Young[1,2], Stephen Thorpe[1], Robert W. Style[3], Eric R. Dufresne[3], and Philipp Kukura[*]

[#]Equal contribution

[1]Physical and Theoretical Chemistry Laboratory, Department of Chemistry, University of Oxford, South Parks Road, Oxford OX1 3TA, UK

[2]Current address: Refeyn Ltd, 1 Electric Avenue, Ferry Hinksey Road, Oxford OX2 0BY, UK

[3]ETH Zürich, 8092 Zürich, Switzerland

[*]To whom correspondence should be addressed: philipp.kukura@chem.ox.ac.uk

*Abstract*

Single particle tracking has found broad applications in the life and physical sciences, enabling the observation and characterisation of nano- and microscopic motion. Fluorescence-based approaches are ideally suited for high-background environments, such as tracking lipids or proteins in or on cells, due to superior background rejection. Scattering-based detection is preferable when localisation precision and imaging speed are paramount due to the in principle infinite photon budget. Here, we show that micromirror-based total internal reflection dark field microscopy enables background suppression previously only reported for interferometric scattering microscopy, resulting in nm localisation precision at 6 μs exposure time for 20 nm gold nanoparticles with a 25 x 25 μm$^2$ field of view. We demonstrate the capabilities of our implementation by characterizing sub-nm deterministic flows of 20 nm gold nanoparticles at liquid-liquid interfaces. Our results approach the optimal combination of background suppression, localisation precision and temporal resolution achievable with pure scattering-based imaging and tracking of nanoparticles at regular interfaces.

*Introduction*

The ability to follow the motion of nanoscopic objects fundamentally relies on their detection and the precision with which we can determine the position of the object of interest for a given integration time[1]. This localisation precision is important for defining the smallest measurable real motion, as opposed to positional fluctuations arising from fitting a signal with finite noise[2], while the integration time determines the type of motion that can be observed, in an extreme reaching the ballistic limit[3,4]. While fluorescence-based approaches have many advantages in the context of specificity and background suppression, they face limitations at the extremes of localisation precision and temporal resolution when using small labels. For single emitters, the fundamental photophysics limits photon emission rates to about 1 MHz if trajectories of appreciable length are to be recorded (>> ms). Coupled with limited detection efficiencies of even the best microscopes, detecting even 1 photon/µs is therefore a formidable task, although limitations these have been mitigated recently by the introduction of MINFLUX[5]. Larger objects or multi-emitter systems can be used[6], although for labels on the order of 40 nm or more one begins to resolve label motion in addition to the motion of the labelled object, thus requiring temporal averaging back to the ms regime in order to maintain the desired localisation precision[7]. An additional limitation for fluorescence imaging arises from photobleaching, which limits the observation time. The latter decreases with increasing illumination power, making trajectories shorter the faster and more accurate the measurement that is required.

One solution to these limitations involves the use of nanoscopic scatterers, which possess greater photostability compared to fluorescent labels. Metallic nanoparticles are ideally suited as scattering labels in the visible regime because they exhibit a plasmon resonance, where their polarizability and thus scattering cross-sections are significantly enhanced[8]. With such labels and improvements to microscope design, nm-precise tracking on the µs timescale has become routine reaching 1.3 nm localisation precision at 55 µs temporal resolution with 40 nm gold nanoparticles (AuNPs)[7,9]. Background suppression using dark-field imaging can in principle be improved to detect even single molecule fluorescence[10], although the more traditional objective-type approaches have struggled with sufficient background suppression to enable high localisation precision with 20 nm labels[11]. Interferometric scattering microscopy (iSCAT) enables observation of even smaller labels, for example reaching 2 nm localisation precision with 2 µs temporal resolution with 20 nm diameter AuNPs[12].

Despite the unique capabilities of iSCAT[13,14], objective-type dark-field imaging has some potential advantages in the context of single particle tracking: 1. Sample illumination is limited to the first few hundred nm near the glass-water interface, reducing both the power load on the bulk sample and background features from spurious scatterers above the interface. 2. High-speed tracking (<ms) with iSCAT requires a stationary illumination beam, which usually limits the field of view to about a <10 µm FWHM if interference fringes caused by focussing the illumination beam into the objective are to be avoided[15]. 3. Objective-type total internal reflection illumination enables trivial combination with total-internal reflection fluorescence imaging as the illumination beam is already provided. 4. The linear scaling of the detected signal for iSCAT with the volume of the scatterer compared to a squared dependence for dark-field detection also implies that sample-specific background that is smaller than the signal of interest plays a more significant role in iSCAT than in dark-field imaging[16]. Conversely, iSCAT is intrinsically more suited to simultaneously imaging a larger range of particle sizes for a given dynamic range in detection.

The last point is illustrated by considering an iSCAT (**Fig. 1a**) image of 20 nm AuNPs on microscope cover glass immersed in water (**Fig. 1b**). Individual AuNPs appear as dark spots on top of a speckle-like background, with a signal-to-background ratio of ~10 (**Fig. 1c**), irrespective of the experimental implementation[12,17,18]. The speckle-like background likely stems from surface imperfections of microscope cover glass, as illustrated when using atomically flat mica surfaces[19] or refractive index matching[20]. While this background can be subtracted computationally[21], it limits the visibility of small particles, and can cause artefacts in localisation. A dark-field image of the same sample could thus, at least in principle, exhibit a 100:1 signal-to-background ratio because of the difference in signal scaling, assuming that spurious background from the illumination can be suppressed to the level of the coverglass-induced scattering.

*Results*

Background light in dark-field microscopy can be generally attributed to two sources: spurious reflections from optics and scattering from or near the illuminated surface itself. (**Fig. S1b**) The achievable signal-to-background ratio in iSCAT (**Fig. 1b**) indicates that the latter can be suppressed significantly below the signal levels from 20 nm gold nanoparticles. Given the extreme levels of background rejection achieved previously by a beam stop placed closely to the entrance pupil[10], we reasoned that using two small mirrors to couple illumination light in and out of the objective in an objective-type TIR arrangement as previously implemented[22,23] (**Fig. 1d and Fig. S1a**), should be able to achieve similar levels of background suppression upon optimisation of the mirror sizes and position. In this experimental arrangement, the main source of background light must thus originate from reflections caused by the lenses contained within the high numerical aperture objective, which could be captured efficiently by mirrors placed near the entrance pupil both for the incident and reflected beam.

A dark-field image of 20 nm diameter gold nanoparticles on a standard glass coverslip immersed in water acquired with a setup using 3 mm diameter mirrors setup exhibited much lower background compared to the nanoparticle signal (**Fig. 1e**) than in the corresponding iSCAT image (**Fig. 1b**). The signal to background ratio now approaches 100:1 as expected both when viewed for individual particles (**Fig. 1f**) and across the field of view (**Fig. 1f inset**). The limiting background is likely caused by coverslip surface roughness as for iSCAT, since it translates when the sample is translated laterally. The key trade-off is between mirror size and position to reduce stray light while minimising loss of scattered light caused by partial blockage of the objective back aperture. To quantify this relationship, we calculated the scattering pattern of a nanoparticle at the glass/water interface at the exit pupil of the objective for s-polarised illumination (**Fig. 2a**)[24–26]. The high emission density at high numerical apertures can be attributed to the refractive index mismatch between glass and water, and illustrates the advantage of using two small mirrors for coupling illumination light compared to a perforated mirror as described previously[11] in terms of collection efficiency. For instance, a mirror with a central aperture of 4 mm diameter combined with an objective with an 8.5 mm diameter entrance pupil would transmit only 15% of collected light.

We then investigated the transmission of a dual mirror approach as a function of both mirror size and position. For the 60x oil immersion objective we use here, a 3 mm diameter mirror enables >65% transmission of scattered light irrespective of the mirror position (**Fig. 2b**). We placed our micromirrors within 5 mm of the entrance pupil along the optical axis, which was the closest practical distance achievable, and at a center-to-center separation of 8.3 mm centred on the entrance pupil (see Methods). For this arrangement, we expect 94% of scattered light collected by the objective to reach the tube lens and ultimately the camera, enabling highly efficient collection. We found excellent agreement between shot noise limited and experimentally determined performance (**Fig. 2c**), evaluated by quantifying interparticle distances (see methods). Detecting ~900 photons at 250 µs exposure time agrees well with the estimated detection efficiency of our setup (40%), the incident power density, and the scattering cross section of 20 nm diameter AuNPs in water (~3 nm$^2$)[27]. We verified the single particle nature of our data by confirming the $V^2$ scaling of the scattering intensity for nominally 20, 30 and 40 nm diameter AuNPs (**Fig. 2d**).

To evaluate the performance of our microscope in a challenging single-particle tracking context, we chose nanoparticles at fluid interfaces, which are an attractive experimental system for studying fundamental questions in condensed matter physics such as the structure and dynamics of confined fluids and phase behaviour of 2D systems. In addition, nanoparticles at interfaces have various technological applications including emulsification, assembly of nanocomposites with tuneable properties and construction of nanostructured membranes for filtration[28–31]. The phenomenon of solid particles adsorbing to soft deformable interfaces, thereby providing them with resistance against coalescence, is well known[32,33], and occurs as a result of the reduction in free energy with respect to the bare interface, as in the case of molecular surfactants. The gain in free energy scales with particle size[34], and thus the stability of nano-and micro-sized particles at the interface increases compared to molecular surfactants. Microparticles are adsorbed irreversibly to interfaces with desorption energies of $\sim 10^8 k_B T$ (the thermal energy). Nanoparticles experience much lower desorption energies of 5-100 $k_B T$ depending on the size and surface chemistry of the particles. The weaker stability of nanoparticles at the interfaces results in a much richer observable energy surface[35], with short-lived metastable states. Imaging at low frame rates of 50 Hz, with exposure times on the ms time scale, enables the observation of time-averaged behaviour. Imaging at much faster frame rates (10-100 kHz) without compromising on the precision opens the possibility to resolve and quantify the more complex potential energy landscapes of these systems.

We thus dispersed AuNPs and silver nanoparticles (AgNPs) at the interface between water and high refractive index (n=1.523) oil (**Fig. 3a**). The high density and refractive index of the oil compared with the microscope cover glass allowed for total internal reflection to occur at the liquid-liquid interface. To enable imaging at high frame rates with nm-precision, while ensuring minimal sample heating, we illuminated in a stroboscopic fashion with a 3.5-5% duty ratio. Each cycle included a dark period of several milliseconds, and an illuminated period of approximately 1 ms, during which one movie of nanoparticle dynamics at the interface was recorded (See methods for more details). In this way, we could generate high contrast dark field images even at very short exposure times (**Fig. 3b**), enabling single particle tracking of tens of particles with simultaneous ~1 nm spatial precision and 10 μs temporal resolution. The particles exhibited differences in their detected intensities (**Fig. 3b** and **Fig. S3**) due to size variations and the strong $a^6$ dependence of the scattering intensity (See methods for more details). Individual trajectories of 20 nm diameter AgNPs showed diffusive motion, in contrast to immobilized AgNPs (**Fig. 3c, d**), where we can eliminate any influence from vibrations (**Fig. S4**). The nanoparticle motion appeared diffusive, evidenced by the linear increase of the mean-squared displacement (MSD) with the lag-time, τ, as expected for conventional single particle diffusion (**Fig.**

3e). In addition, we found no signatures of structuring of the particles at the oil-water interface, evident from the pair correlation function (**Fig. S5**).

To further characterize the nature of the nanoparticle motion at the interface with short lag-times, we quantified the spatial dependence of single particle displacements (**Fig. 4a**). Although the particle trajectories and MSD appear diffusive, the vector field at short lag-times (τ = 10 μs) revealed an outward flow pattern with drift displacements smaller than 0.5 nm, corresponding to a drift velocity of 50 μm/s (**Fig. 4b**). While conventional Brownian motion would result in Gaussian distributions centred on zero, these velocity distributions show two well-resolved peaks at non-zero speeds. Interestingly, this behaviour persists for different particle concentrations, with faster speeds at higher particle concentrations (**Fig. S6**).

Deterministic flows at fluid-fluid interfaces[36] can be driven by non-uniform surface tensions. This effect is termed Marangoni flow, and typically arises from temperature or surfactant concentration gradients. Here, the absorption of the laser illumination by AgNPs drives local heating. In the Supplement, we show that a temperature difference as small as 1 mK degrees over the 10 micron field-of-view is sufficient to drive the observed flow. This hypothesis is supported by the observation that the flow speed increases at higher particle concentrations (**Fig. S6**).

To further demonstrate that this behaviour is an interfacial effect rather than a bulk one, we analysed the correlated motion of pairs of particles, which allows us to observe the fluid-mediated interactions between particles. For that purpose, we quantified the ensemble-averaged longitudinal ( ∥ ) displacement correlations of particle pairs as a function of inter-particle distance $r$ and lag time $\tau$[37]:

$$\langle \Delta x_1 \Delta x_2 \rangle_\parallel \equiv \langle \Delta x_i^\parallel(t,\tau) \Delta x_j^\parallel(t,\tau) \delta\left(r - R_{ij}(t)\right) \rangle$$

where $\Delta x_i^\parallel(t, \tau)$ is the displacement of particle $i$ between time $t$ and $t + \tau$, projected parallel to the line connecting the pair, and $R_{ij}(t)$ is the pair separation at time $t$ (see also inset of **Fig. 4c**). The angular brackets represent averages over all times $t$ and particle pairs $i \neq j$. In systems where the momentum can travel in three dimensions, these correlations should decay as $\sim 1/r$. For diffusive motion these correlations are expected to increase linearly with $\tau$, similar to the MSD. Intriguingly, at short time intervals, the longitudinal displacement correlations scale as $\sim \ln(1/r)$, characteristic of interface-dominated behaviour[38–40] (**Fig. 4c**), where momentum travels only in two dimensions. Re-plotting the displacement correlation for each $r$ as a function of $\tau$ shows super-diffusive scaling, in the vicinity of the ballistic limit, which scales with $\sim \tau^2$ (**Fig. 4d**). This reflects the dominance of deterministic outward motion (**Fig. 4a**). The plots in **Fig. 4d** are normalized by their initial value (i.e., value at the minimum lag-time, τ$_{min}$) to simplify presentation. Similar behaviour was observed for 20 and 30 nm AuNPs (**Fig. S7**).

Operating with stroboscopic illumination allowed us to study the particle motion at both short (10-100 µs) and long (40-400 ms) lag-times, simultaneously. In the long lag-time movies, the laser illuminated the sample for less than 5% of the total lag-time, allowing the system to equilibrate and any local heating to dissipate to the bulk. Therefore, at long lag-times we observed no apparent flows (**Fig. 4e**), with drift velocities less than 1 µm/s, distributed around zero (**Fig. 4f**). This indicates that the system had sufficient time to equilibrate during the dark period. In accordance, the displacement correlation curves exhibit normal diffusive behaviour with a $r^{-1}$ decay (**Fig. 4g**) and linear time dependence (**Fig. 4h**). These results indicate that the viscous response at the interface is mediated by the two bulk fluids (oil and water) rather than the interface, as anticipated (see supplement for a full derivation of the expected viscous response).

We further analysed the displacement correlations at short lag-times (**Fig. 4c**) to separate the deterministic behaviour, which scales as $\sim \tau^2$, from the diffusive (viscous) behaviour, which scales as $\sim \tau$ (**Fig. S8,** supplement for analysis details). Extraction of the correlated diffusive coefficients provides a mean to estimate the effective viscosity at the interface, related to the flows at and close to the interface. For large nanoparticle separations, the leading term of this effective viscosity should be only influenced by the properties of the two bulk fluids (and not the interface), with each fluid contributing equally, i.e., insensitive to the exact lateral position of the particle within the interface. Therefore, the effective viscosity is expected to be the average viscosities of the two bulk fluids (see methods for the full derivation). This viscosity is different from the local viscosity extracted from self-diffusion of particles, which is sensitive to the exact position of the particles at the interface (i.e., their immersion depth or three-phase contract angle). The effective viscosity values were approximately the average between the oil and water viscosities (0.044 and 0.00095 $Pa \cdot s$, respectively), corresponding $0.0241 \pm 0.003$ (20 nm AgNPs), $0.0227 \pm 0.0034$ (20 AuNPs) and $0.0244 \pm 0.004$ (30 AuNPs) $Pa \cdot s$. These values agree with the theoretical prediction based on the Green functions at fluid-fluid interfaces (see methods and **Fig. S9**).

*Discussion*

Our results demonstrate that: (i) Background suppression in micromirror-based TIR can be comparable to that achievable in iSCAT, reaching the limits set by microscope cover glass. (ii) Such an experimental approach can be advantageous in reaching high levels of efficiency in scenarios where a scattering label of interest produces stronger image contrast than imaging background. (iii) The combination of localisation precision and temporal resolution is some of the highest reported to date for the size of scattering labels used and approaches the optimum achievable imaging performance for nanoparticle tracking in the absence of additional measures aimed at enhancing the illumination field. (iv) These advantages can be achieved in a package that enables facile multicolour illumination and simultaneous scattering and fluorescence imaging, while taking advantage of background suppression enabled by evanescent illumination.

To demonstrate the capabilities of our microscope, we selected a system of nanoparticles at an oil-water interface. We were able to record the motion of AuNPs and AgNPs on the 10-100 µs timescale (high frame rates of 10-100 kHz), and a large field of view of tens µm$^2$, thus capturing the trajectories of tens of nanoparticles simultaneously. Moreover, our stroboscopic illumination approach allowed us to also simultaneously record particle motion at longer time intervals of tens of ms. At short time intervals the averaged MSD suggests particles exhibit diffusive behaviour. However, further analysis of the particles' motion revealed spatial correlations and a deterministic motion which emerged on top of the diffusive one. We linked the latter behaviour to thermal Marangoni flows, which decay within a few milliseconds once the illumination is turned off and therefore are not present at long-time intervals. In addition, we could separate the diffusive and deterministic contributions of the displacement correlation behaviour. The diffusive contribution to the correlations at large particle separations is in excellent agreement with theoretical predictions, described in the Supplement. According to these calculations, the effective viscosity of this mode should be equal to the average viscosity of the two phases. The effective viscosity of this mode was found to be $0.0237 \pm 0.0035$ Pa $\cdot$ s, while the average viscosity of the two phases is $0.0225$ Pa $\cdot$ s.

Taken together, these results demonstrate that micromirror darkfield microscopy can achieve background suppression towards the limits previously reported with iSCAT and can be used for ultra-high performance particle imaging and tracking. For low background tracking applications, such as those on microscope cover glass including supported lipid bilayers or molecular motors, where scattering labels producing signals comparable to 20 nm AuNPs can be used, our approach may prove to be the preferred choice due to its combination of detection efficiency, background suppression, field of view and achievable imaging speed.

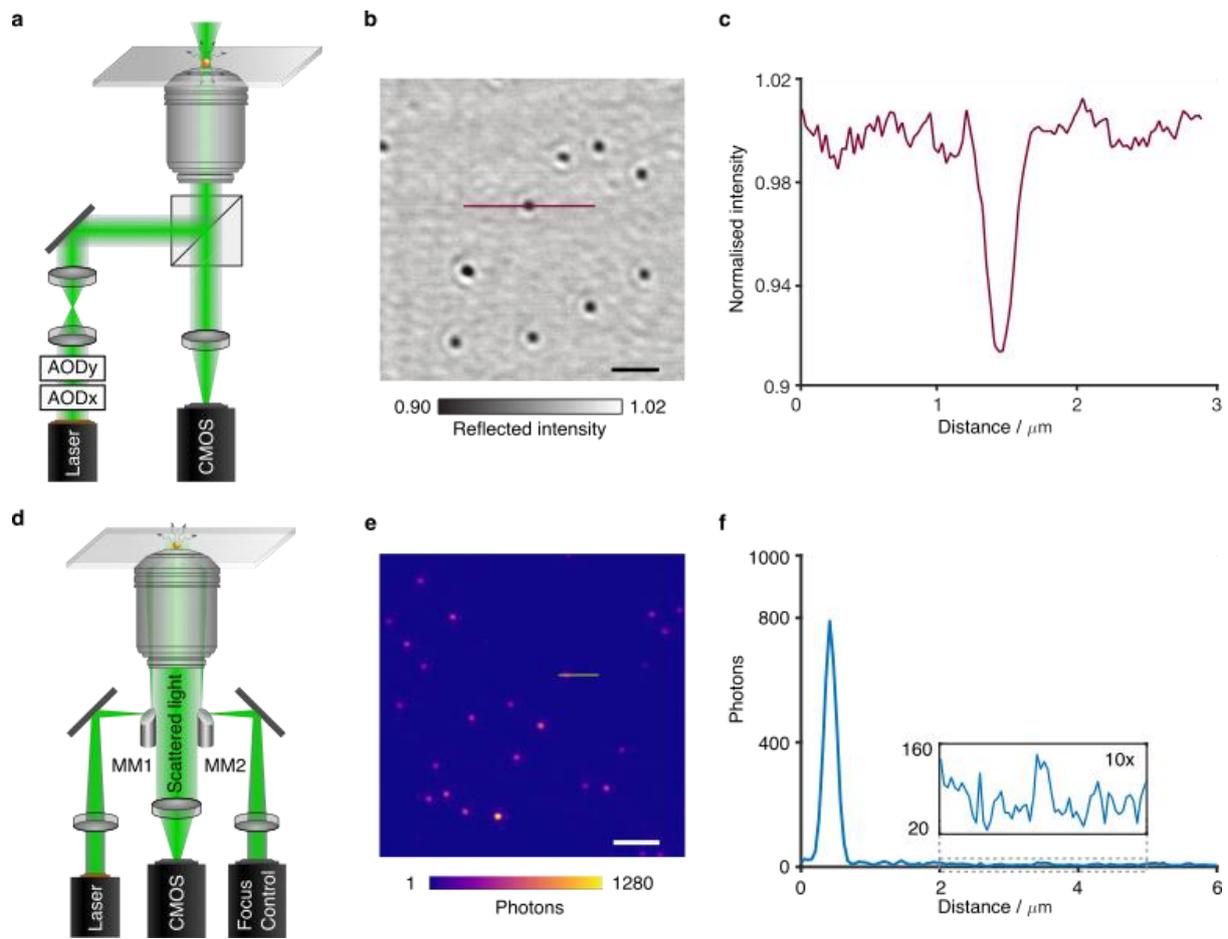

**Figure 1: Comparison of interferometric scattering (iSCAT) and micro-mirror based total internal reflection scattering microscopy. a**, Schematic illustration of an AOD-based iSCAT microscope. **b**, Typical iSCAT image of 20 nm AuNPs immobilized on a cover glass surface immersed in water. Scale bar: 1 µm. **c**, Line cut across the particle in **b**. **d**, Objective-type TIR using two micro-mirrors. **e**, Respective image of 20 nm AuNPs immobilised on cover glass Scale bar: 2 µm. **f**, Linecut of a single particle in **e**, including a zoom of the imaging background. AOD: Acousto-optic deflectors, T: Telecentric lens system, BS: Beam splitter, L: Lens, MM: Micro-mirror.

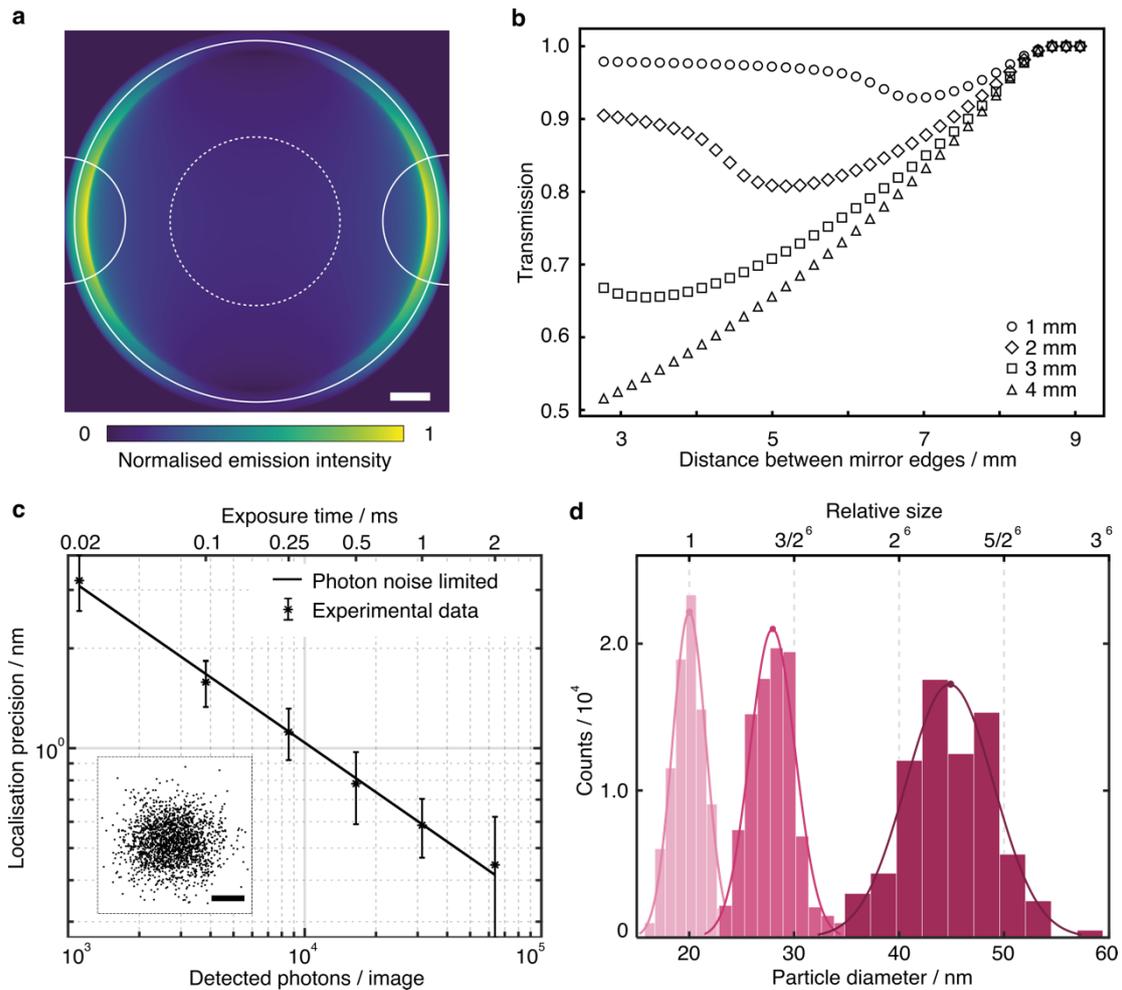

**Figure 2. Performance of the micromirror-TIRS microscope. a**, Back focal plane emission pattern of a single emitter at a glass (n= 1.52) – water (n = 1.33) interface. The objective pupil + ring (NA = 1.42, Back aperture = 8.52 mm) and micro-mirrors are indicated by white solid lines. The aperture of a 4 mm diameter perforated mirror is shown by white dashed lines. **b**, Transmission as a function of mirror size and separation for the objective and mirrors shown in **a**. **c**, Theoretical shot-noise limited behaviour (line) and experimental localisation precision measured by varying the camera exposure time at 2.7 kW/cm$^2$ illumination intensity. **d**, Measured relative scattering signal for different nanoparticle sizes and comparison with D$^6$ scaling.

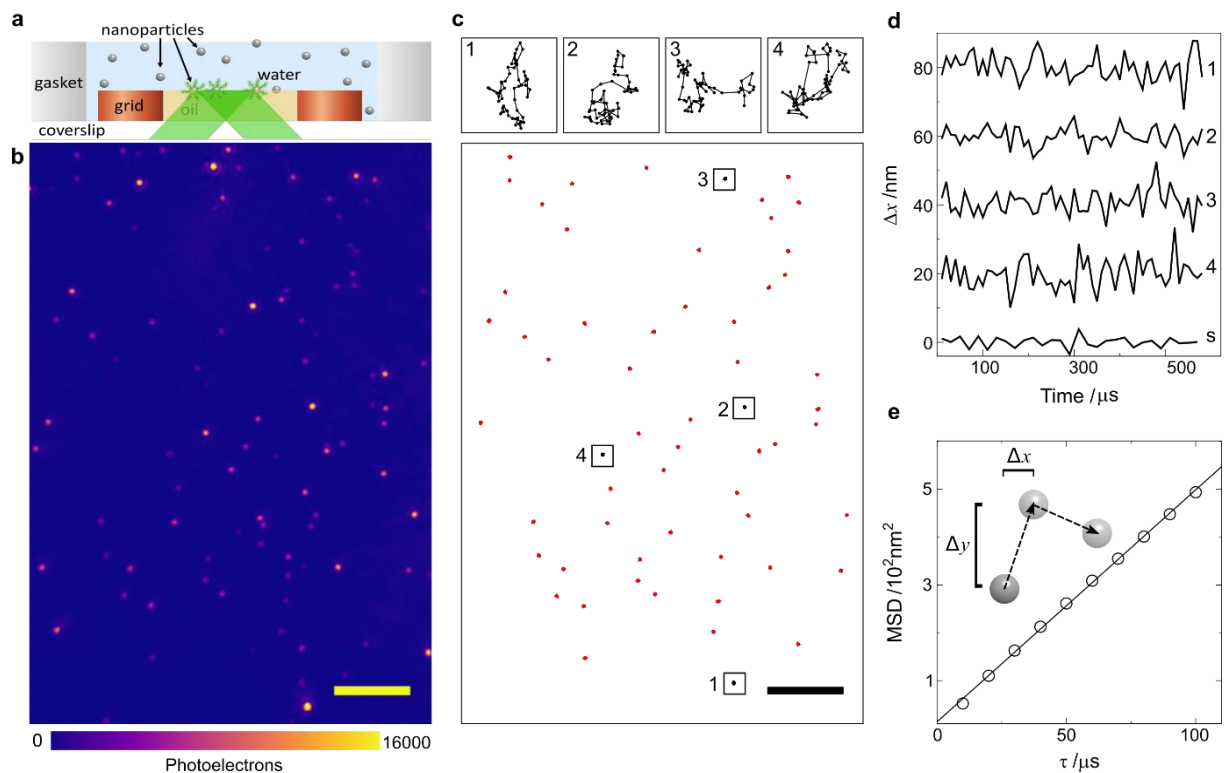

**Figure 3. High speed diffusion of metallic nanoparticles at oil-water interfaces. a**, Schematic illustration of the experimental system; nanoparticles (NPs) are dispersed at an oil-water interface. **b**, Representative image of 20 nm AgNPs diffusing at the interface. Scale bar: 4 µm. **c**, Trajectories of 20 nm AgNPs acquired at 100 kHz during the illumination period (in total 60 frames), and zoom of 4 trajectories. Scale bar: 4 µm. **d**, Step-sizes along the *x*-axis, $\Delta x$ for the 4 zoomed trajectories in **c.** The $\Delta x$ traces are shifted vertically for clarity. The bottom trace corresponds to an immobilised particle on glass (s). **e**, Ensemble and time-averaged mean square displacement (MSD) of 20 nm AgNPs at an oil-water interface vs lag-time $\tau$ (data taken at 100,000 Hz). The red line represents a linear fit, MSD = $4D_s\tau$, to the data. Inset: illustration of particle diffusion.

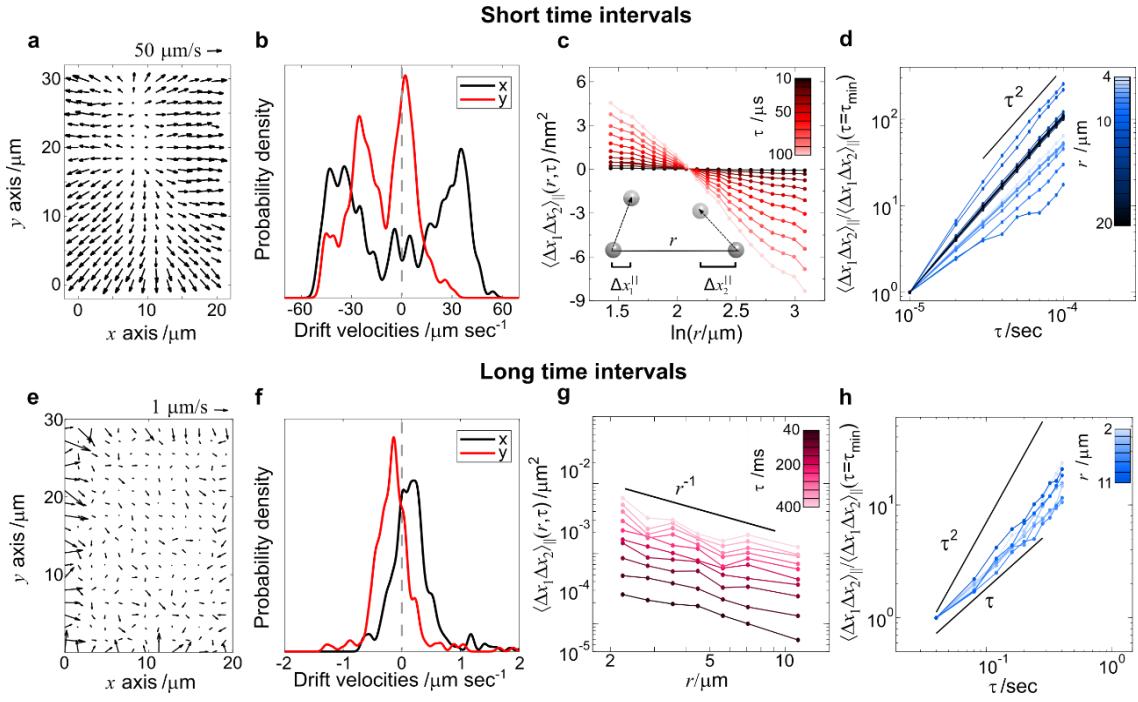

**Figure 4. Small-scale effects of nanoparticle diffusion at the oil-water interface. a-d**, Short time intervals: **a**, Displacement field of 20 nm AgNPs at 100 kHz. Scale bar 50 µm/s. **b**, Probability density of the average velocities from **a** along x (black) and y (red). **c-d**, Two-point displacement correlation at oil-water interfaces for 20 nm AgNPs, imaged at 100 kHz. **c**, Longitudinal displacement correlation as a function of the logarithm particle separation, ln($r$), for 10 different lag-times, τ = 10-100 µs. **d**, Replotting the displacement correlation for each $r$ as a function of τ. Plots are normalized to the first data point (i.e., value at the minimum lag-time, $τ_{min}$) for clarity. **e-h**, Long-time intervals: **e**, Displacement field of 20 nm AgNPs at 50 Hz. Scale bar 1 µm/s. **f**, Probability density of the average velocities from **e** along x (black) and y (red). **g**, Longitudinal displacement correlation as a function of $r$, for different lag-times between τ = 0.04 – 0.4 sec. **h**, Replotting the displacement correlation for each $r$ as a function of τ. Plots are normalized as for **d**.

**Materials and Methods**

*Materials*

The immersion oil used to set up the oil-water interface is an immersion liquid (Cargille Laboratories, Cat.# 1160), where the oil viscosity and density according to the manufacturer are $\eta_o = 0.044$ Pa·s and $\rho = 1.1$ g/cm$^3$, respectively. Particulates were reduced before each experiment by centrifugation at 13,500 rpm for at least half an hour. We used gold (AuNP) or silver (AgNP) nanoparticles with diameters of 21.1 ± 3.1 (AgNPs), 19.1 ± 2.4 (AuNPs) or 29.2 ± 4.2 nm (AuNPs), all purchased from nanoComposix with surface passivation of polyvinylpyrrolidone (PVP) to prevent particle aggregation (Lots# ECP1386 - 20 nm AuNPs, SCM0057 20 nm AgNPs, ECP1308 30 nm AuNPs). All NPs are purchased as a suspension in water without any additional stabilizers. Water used in our experiments, both for cleaning and dilution of particles, was Millipore filtered with a resistance of 18.2 MΩ·cm. The oil-water surface tension was 31 ± 1 mN/m, measured using the hanging drop method.

*Microscope alignment*

The construction of the TIRSM setup (**Fig. S1**) comprises three stages in sequence: alignment of the illumination path, alignment of the detection arm and optimisation of background rejection. The entire microscope is built on a large and heavy breadboard and then placed on a vibration isolation optical table spaced with air-filled inner tubes to minimise vibrations from the environment being coupled into the microscope (**Fig. S4**). The setup is fully enclosed into compartments to eliminate background light and instabilities induced by air flow.

Three laser diodes are selected as illumination sources, with emissions wavelengths at 445 nm (LDM-445-6000-C), 520 nm (LDM-520-1000-C) and 635 nm (LDM-635-200-C, Lasertack) used for various applications. For example, the green laser is mostly used for high-speed tracking of gold nanoparticles due to their plasmon resonance. Illumination beams are combined in a plane parallel to the breadboard via dichroic mirrors before being focused near the incident micro-mirror. We use an iris to hard aperture the beam profile to a diameter of approximately 5 mm prior to the focussing lens.

We glued both micro-mirrors on 3D printed mirror holders and attached them to high stability kinematic mirror mounts to enable tilt control. The first micro-mirror is placed into the illumination path and its height is adjusted to ensure the illumination beam strikes the area close to the tip of the micro-mirror. This allows for maximal mirror separation in the following step to minimise loss of scattered light. An iris mounted at the same height behind the incident micro-mirror serves as a reference for future alignment of the second micro-mirror. We adjusted the incident micro-mirror such that the illumination beam travels straight upwards and through the centre of the objective. To achieve this, a lens tube with two irises mounted on both sides is screwed into the objective thread. When the beam travels through the centre of both irises, we marked a target in its position on the ceiling which serves as a reference for later beam positioning. Subsequently, the objective is inserted back and translated perpendicular to the incident beam until the beam hits the target on the ceiling. We chose a 200 mm focal length lens and positioned it in the excitation path without introducing beam displacements relative to the ceiling target. It is moved back and forth until the beam size projected onto the ceiling is the smallest to ensure that the lens focuses the beam into the back focal plane (BFP) of the objective.

To carry out the alignment of the detection path, the rod mirror is translated out of the imaging path and a 45° mirror is placed underneath the objective to collect scattered light. An additional alignment laser is applied from the camera side and collimated to travel straight in a plane parallel to the optical table towards the 45° mirror. The objective was temporarily removed and replaced by a flat mirror, adjusting the 45° mirror until the reflected beam from the flat mirror overlaps with the excitation beam. After that, we explored its final position by moving the 45° mirror to ensure the reflected beam on the ceiling exhibits no displacement relative to the target in the absence of the flat mirror. To add the camera, the objective was inserted back but with a ring-actuated iris diaphragm attached to its back aperture. We applied a diffuse light source close to the objective front lens and adjusted the diameter of the iris to achieve a few mm diameter beam spot emerging from the objective. We then placed the camera in place ensuring it is perpendicular to the imaging path and the beam spot is aimed precisely at the middle of the chip. Next, a 200 mm imaging lens is introduced at the appropriate focal distance away from the camera without causing beam displacements on the camera chip. The distance between the lens and the camera sensor is deemed correct when the beam profile has the largest intensity and smallest area for an incident collimated beam.

Finally, with the illumination and detection path aligned, we investigated the intensity of dark background images with only scattered light from clean cover glass surface roughness and stray light induced by the refractive index mismatch at the interface to explore the optimal position and separation of mirrors. Starting by translating the incident micro-mirror towards the periphery of the objective in the direction parallel to the beam path, a strong returning beam appears upon total internal reflection. We stopped before the totally reflected beam distorted and moved the second micro-mirror accordingly to pick up the returning beam. After that, we repeated the following steps while constantly checking the imaging background caused by cover glass roughness. Background suppression was deemed optimised when the background minimised while the signal to background ratio of single 20 nm AuNPs is the highest:

a. Move micro-mirrors in the direction perpendicular to the beam path until the lowest background intensity is observed.

b. Move the first micro-mirror backwards in 0.2 mm steps and consequently the second micro-mirror, and repeat step a.

c. Repeat step b at least 10 times and retreat micro-mirrors where necessary to explore the lowest background noise level.

d. Fix the micro-mirrors at the position with simultaneously lowest background noise and largest mirror separation.

*Focus control*

The focus plane was stabilised by recording the position of the total internally reflected beam based on the principle that drifts in the displacement between the sample and the objective translate into a translation of the reflected beam[18] (**Fig. S2**). We chose an infrared laser (λ = 830 nm) to minimise sample perturbation. It was aligned and coupled into the illumination arm with the aid of two mirrors. When TIR appears at the interface, the reflected beam is redirected by the second micro-mirror onto a focus control camera (PointGrey, Firefly USB 2). We used a cylindrical lens (CyL) to stretch the beam on the camera perpendicular to the translation for improved determination of its centre position. It is essential to place the cylindrical lens away from the focal distance relative to the camera chip in order to improve the sensitivity of the imaged beam to lateral displacements, and thus focus. Focus drift associated with the beam displacement on the camera is extracted by calculating the centre of mass of the linear beam profile after summing over all camera rows. To eliminate contributions from other illumination wavelengths, a long pass filter is applied in front of the camera. Note that high pointing stability of the illumination beam is required since this approach is sensitive to beam profile and intensity fluctuations which in turn may lead to unwanted beam displacements hence deceptive changes in focus. We generally switched the focus laser off during data acquisition.

*Calculation of scattered light transmission*

The electric field intensity *I* for a single point scatterer at a glass water interface was estimated using the Weyl representation[24]. The NA is calculated along a 2D numerical aperture co-ordinate, *NA$_{map}$*, in the back aperture plane up to *N$_{immersion}$*:

$$NA_{map}(x,y) = NA_{immersion} \frac{\sqrt{x^2 + y^2}}{x_{max}}$$

The emitter was placed at the glass surface, z = 0, the dipole oriented parallel to the glass, ϑ = 90, and along the incident polarisation, which is perpendicular to the plane of incidence (Φ = 90). The intensity including the NA shaping of the objective is:

$$I_{objective} = I M_{objective}$$

Where M$_{objective}$ is:

$$M_{objective}(x,y) = \begin{cases} 0 & (NA_{map} > NA_{objective}) \\ 1 & (NA_{map} < NA_{objective}) \end{cases}$$

The pixel size is calibrated by dividing the sum of NA$_{map}$ across the centre of the image by the objective back aperture. The NA of a single micromirror *NA$_{mirror}$* is:

$$NA_{mirror} = NA_{objective} \frac{D}{B}$$

where *D* is the diameter of the mirror and *B* is the objective back aperture diameter mask, Bap = 8.52 mm.

The intensity including NA shaping of the two micromirrors is:

$$I_{mirror} = I_{objective}(M_{mirror}(-x_{dist})M_{mirror}(+x_{dist}))$$

Where M$_{mirror}$ is shifted in *x* by x$_{dist}$, the half radial distance between the two micromirrors. The NA shaped intensity was evaluated from the minimum physical distance between the mirrors in the real system (2.6 mm) to a position entirely outside of the objective back aperture diameter (9.06 mm) for mirrors of 1 to 4 mm diameter. The measured transmission of scattered light is:

$$T_r = \frac{\sum I_{mirror}}{\sum I_{objective}}$$

*Data acquisition and set-up adaptation:*

For the interface experiments, the light scattered by NPs was collected by the objective and imaged onto a high-speed CMOS camera (Photron SA-X2). The final magnification of our TIR scattering microscope is 250x, resulting in 80 nm/camera pixel for the camera used. The detected intensity varied substantially from particle to particle (see **Fig. 3b** of the main text and **Fig. S3**, black line), due to the strong $a^6$ dependence of the scattering intensity to variation in particle size. The size variation of our NPs is 15%, resulting in broad distributions of scattering intensity. To confirm that the origin of our intensity variations is predominantly caused by the non-uniform size distribution of our particles, we generated the expected intensity distribution for an array of particles with diameters corresponding to the known size distribution (**Fig. S3**, green line). The expected intensities for our NPs have a distribution with a heavy tail of large intensities. The camera has an upper limit (or maximum number) of detected photo-electrons for each pixel (full well capacity) of 16,000 photoelectrons, and therefore all particles with higher intensities will saturate the camera, exhibiting the same peak intensity. To account for the saturation effect, all particles with peak intensities above this threshold (16,000) were given a constant peak-intensity value of 16,000. The distribution of these 'corrected' (integrated) intensities is plotted in **Fig. S3** (red line), with an artificial peak appearing at larger intensities. This second peak appears also in the experimental intensity distribution and is a result of the effects discussed above (scattering cross-section and camera saturation). This comparison confirms that the dominant contribution to scattering intensity variations is the underlying particle size distribution, rather than aggregation, for example.

To enable imaging at high frame rates with nm-precision without excessive heating, we illuminate in a stroboscopic fashion with a 3.5-5% duty ratio, and cycle length of 20-40 ms. Pulses were generated by a function generator, which simultaneously triggered the light source and camera to synchronise illumination and acquisition. Within the pulse duration, which varied between 500-1000 μs, the power density at the sample was ~ 30 kW/cm$^2$, while the integrated power density over the entire cycle was ~ 2 kW/cm$^2$. Every pulse produced a movie of 10−60 images depending on the frame rate used, between 10,000-100,000 frame per second (fps), and a sequence of ~4000 movies from consecutive pulses were captured. A typical dark-field image is shown in **Fig. 3b**. Nanoparticles were localised by an automated image analysis software, which included a bandpass filter to remove noise at high and low frequencies[41], an intensity threshold to estimate particle candidates, followed by 2D Gaussian fitting. Particle tracking was performed by correlating particle positions in consecutive frames[41]. Trajectories of NPs at short lag-times were obtained within individual pulses, while trajectories at long lag-times were generated by tracking the motion of particles between consecutive pulses. This analysis allowed us to simultaneously explore the dynamics of particles on vastly different timescales, of 10-100 μs and 10-100 ms.

To verify that we track only interfacial particles, i.e. particles that are absorbed to the interface, we filtered particles by several criteria[42]. First, since we use TIR illumination, we are only able to see particles which are located within a few 100 nm above the fluid-interface. This feature already eliminates most particles suspended at the bulk of the water phase. In addition, only particles that could be tracked for at least 10 consecutive movies were considered for the remaining analysis, which avoids NPs that are floating just above the interface. Finally, exploiting the large difference in diffusivity between particles at the interface and in the water phase (×20 larger diffusivity in bulk water phase), we measured the self diffusion coefficient ($D_s$) for each particle from its mean squared displacement curves (MSD), $D_s = \text{MSD}/4\tau$, where $\tau$ is the lag-time. We then removed particles whose trajectories exhibited an unexpectedly high diffusivity (larger by a factor of 7 from the average value of all particles in the experiment), as these values would correspond to particles which diffuse just above the fluid-interface.

*Microscope coverslip cleaning procedure*

We cleaned borosilicate microscope coverslips (No. 1.5, 24 × 50 mm, VWR) by rinsing them sequentially with water, ethanol, water, isopropanol, water, ethanol and water, followed by drying under a clean stream of nitrogen.

*Preparation of oil-water interfaces*

Our oil-water interfaces are decorated sparsely with metallic nanoparticles. The interface was set up in a similar fashion to recent studies exploiting TEM grids to stabilize hydrophobic phases by capillarity[35,43,44]. Here, we used TEM copper grids with an 1.5 mm aperture (Gilder Grids, cat.# GA1500) to stabilize the immersion liquid on clean coverslips. The TEM grid was confined by a polydimethylsiloxane (PDMS) gasket with an inner diameter of ~4 mm. Subsequently, a 20 μL water droplet containing the NPs (with a concentration in the range of $10^{-13} - 10^{-12}$ M) was placed on top of the oil-filled grid and the coverslip was left on the microscope sample holder for at least 30 mins to equilibrate before acquisition. During this time the NPs settled at the interface. Due to their small surface area, the NPs are expected to settle on the seconds to minutes timescale at their equilibrium position within the interface[45].

**Acknowledgments**

P.K. is supported by an ERC Consolidator grant (PHOTOMASS 819593), S.T. by the Emerson Collective.